\documentclass[numsec,webpdf,modern,large,namedate]{oup-authoring-template} %
\usepackage{subfigure}
\theoremstyle{thmstyleone}%
\theoremstyle{thmstyletwo}%
\theoremstyle{thmstylethree}%

\usepackage{hyperref}
\hypersetup{colorlinks=true, citecolor=blue, urlcolor=blue, linkcolor=blue}
\begin{document}
\journaltitle{Bioinformatics}
\DOI{https://doi.org/xxx}
\copyrightyear{xxxx}
\appnotes{\bf ISMB 2024}

\firstpage{1}

\title[Short Article Title]{RNACG: A Universal RNA Sequence Conditional Generation model based on Flow Matching}

\author[1,2]{Letian Gao\ORCID{0000-0003-1232-8866}}
\author[1,2,$\ast$]{Zhi John Lu\ORCID{0000-0003-3739-8756}}

\authormark{Author Name et al.}

\address[1]{MOE Key Laboratory of Bioinformatics, Center for Synthetic and Systems Biology, School of Life Sciences, Tsinghua University, Beijing 100084, China.}
\address[2]{Institute for Precision Medicine, Tsinghua University, Beijing 100084, China.}

\corresp[$\ast$]{Corresponding author: zhilu@tsinghua.edu.cn}




\abstract{
  \textbf{Motivation}: RNA plays a pivotal role in diverse biological processes, ranging from gene regulation to catalysis. Recent advances in RNA design, such as RfamGen, Ribodiffusion and RDesign, have demonstrated promising results, with successful designs of functional sequences. However, RNA design remains challenging due to the inherent flexibility of RNA molecules and the scarcity of experimental data on tertiary and secondary structures compared to proteins. These limitations highlight the need for a more universal and comprehensive approach to RNA design that integrates diverse annotation information at the sequence level.\\
  To address these challenges, we propose RNACG (RNA Conditional Generator), a universal framework for RNA sequence design based on flow matching. RNACG supports diverse conditional inputs, including structural, functional, and family-specific annotations, and offers a modular design that allows users to customize the encoding network for specific tasks. By unifying sequence generation under a single framework, RNACG enables the integration of multiple RNA design paradigms, from family-specific generation to tertiary structure inverse folding.\\
  \textbf{Results}: We evaluate RNACG on multiple benchmarks, including RNA 3D structure inverse folding, family-specific sequence generation, and 5’UTR translation efficiency prediction. RNACG achieves competitive or state-of-the-art performance across these tasks, demonstrating its versatility and robustness. Specifically, RNACG outperforms existing methods in sequence recovery rate and F1 score while requiring significantly fewer parameters. Our results highlight the potential of RNACG as a powerful tool for RNA sequence design, enabling large-scale simulation experiments and advancing applications in synthetic biology and therapeutics.
}

\maketitle
\section{Introduction}

The design and development of specific biomolecules remain a critical challenge in fields such as synthetic biology and immunology, with RNA design playing an indispensable role. RNA molecules are versatile tools in biology, capable of regulating gene expression, recognizing specific small molecules as ligands, and catalyzing biochemical reactions as ribozymes \citep{ref:selectionfunctionalnucleicacids}. Despite their functional diversity, RNA design has lagged behind protein design due to the inherent flexibility of RNA molecules and the scarcity of experimental data on their tertiary and secondary structures \citep{ref:foldandfind}. While proteins benefit from a wealth of structural data and advanced computational tools like AlphaFold \citep{ref:alphafold3}, RNA design remains a challenging frontier.

Recent advances in RNA design have demonstrated promising results. For example, RfamGen has successfully reconstructed covariance models (CMs) for family-specific sequence generation, enabling the design of functional ribozymes with catalytic activities exceeding those of natural sequences \citep{ref:rfamgen}. Similarly, Ribodiffusion and RDesign have made significant strides in RNA inverse folding, leveraging tertiary structure information to design functional sequences \citep{ref:ribodiffusion,ref:rdesign}. These methods have shown that RNA design can achieve practical utility, particularly in applications such as synthetic riboswitches \citep{ref:riboswitchdesign1,ref:riboswitchdesign2}.

However, RNA design still faces significant challenges. The flexibility of RNA molecules makes it difficult to predict and stabilize their tertiary structures, and experimental data on RNA structures remain limited compared to proteins \citep{ref:bpRNA}. This scarcity of data hinders the development of robust computational models for RNA design. Moreover, existing methods often focus on specific aspects of RNA design, such as secondary structure prediction \citep{ref:mxfold2} or family-specific sequence generation \citep{ref:rfamgen}, without integrating diverse annotation information into a unified framework.

To address these challenges, we propose RNACG (RNA Conditional Generator), a universal framework for RNA sequence design that integrates multiple design paradigms, from family-specific generation to tertiary structure inverse folding. By leveraging flow matching \citep{ref:cfmlipman} and Dirichlet distributions \citep{ref:dirichletfm}, RNACG enables the generation of RNA sequences that satisfy diverse structural and functional constraints. Our approach builds on recent advances in deep learning, such as diffusion models \citep{ref:dit} and geometric learning \citep{ref:se3transformer}, to provide a flexible and scalable solution for RNA design.

In this study, we evaluate RNACG on multiple benchmarks, including RNA 3D structure inverse folding, family-specific sequence generation, and 5'UTR translation efficiency prediction. Our results demonstrate that RNACG achieves competitive or state-of-the-art performance across these tasks, highlighting its potential as a powerful tool for RNA sequence design. By unifying sequence generation under a single framework, RNACG paves the way for more comprehensive and efficient RNA design strategies, with applications in synthetic biology, therapeutics, and beyond.

\section{Methods}

In this section, we provide a detailed description of RNACG (RNA Conditional Generator), including the formulation of RNA design problems in sequence space, the principles of flow matching, the overall architecture of RNACG, and the specific training strategies for different tasks. Our goal is to present a unified framework that integrates diverse RNA design paradigms, from family-specific sequence generation to tertiary structure inverse folding, while maintaining computational efficiency and scalability.

\subsection{Problem formulation}

\subsubsection{Formulation of RNA Design Problems}

Given a target property \( c^* \) (such as a specific structure, RNA type, or function), RNA design aims to identify a function \( f \) that maps the property \( c \) to the set of all RNA sequences \( S \) that exhibit the desired property. Formally, this can be expressed as:

\begin{equation}
\exists f: c \mapsto \{S \in \Sigma^* \mid S \text{ has property } c^*\}
\end{equation}

In most practical scenarios, it is infeasible to find a sequence \( S \) that perfectly satisfies the property \( c^* \). Instead, we rely on a distance metric \( \mathcal{D} \) to measure the discrepancy between the property \( c^* \) of \( S \) and the target property \( c \). The function \( f \) can then be redefined as:

\begin{equation}
\exists f: c \mapsto \{S \in \Sigma^* \mid S \text{ has property } c^* \text{ and } \mathcal{D}(c, c^*) \leq \epsilon\}
\end{equation}

This leads to the following optimization problem:

\begin{equation}
\min_{S \in \Sigma^*} \mathcal{D}(c, c^*)
\end{equation}

An RNA sequence is a string of nucleotides, represented as \( S = \{s_i\}^L = \{A, C, G, U\}^L \), where \( L \) is the length of the sequence. The probability of each nucleotide \( s_i \) is given by:

\begin{equation}
P(s_i \mid \Sigma^*) = \sum_{s_{ki} = s_i; S_k \in \Sigma^*} P(S_k \mid \Sigma^*)
\end{equation}

The probability of a sequence \( S_k \) in the set \( \Sigma^* \) can be expressed as:

\begin{equation}
\label{eq:target_set}
P(S_k \mid \Sigma^*) = \frac{\prod_i P(s_{ki} \mid \Sigma^*)}{\sum_{S_j \in \Sigma^*} \prod_i P(s_{ji} \mid \Sigma^*)}
\end{equation}

Based on the above, the general optimization goal for RNA design problems can be formulated as:

\begin{equation}
\label{eq:RNA_design_goal}
\begin{aligned}
\min \mathbf{E} (\mathcal{D}(c, c^*)) \Leftrightarrow \min & \sum_{S_k \in \Sigma^*} \mathcal{D}(c, c^*) P(S_k \mid \Sigma^*)
\end{aligned}
\end{equation}

\subsubsection{Formulation of RNA Design Model}

According to Eq \eqref{eq:RNA_design_goal}, we need to define the interval \( S \in \Sigma^* \), the distance metric \( \mathcal{D} \), and the probability \( P(s_i) \). 
If the mapping function \( g \) between the property \( c \) and the sequence \( S \) is known, we can use reinforcement learning to optimize the goal. However, in most cases, the mapping function \( g \) is either unknown or inaccurate.

Here, we transform the function \( f \) with parameter \( \theta \) as:

\begin{equation}
  \label{eq:RNA_design_function}
  \exists f_{\theta}: c^* \to S \textrm{ where } S \textrm{ has property } c^*
\end{equation}

The distance metric \( \mathcal{D} \) can be defined as:

\begin{equation}
  \mathcal{D}(c, c^*) = 1 - \mathcal{D}^{'} (f_{\theta}(c^*), S)
\end{equation}

Then, the optimization goal can be formulated as:

\begin{equation}
    \label{eq:RNA_design_goal_2}
    \begin{aligned}
   \theta & =  \arg \min_{\theta} \sum_{S_k \in \Sigma^*} \mathcal{D}(c, c^*) P(S_{k}|\Sigma^*)\\
    & = \arg \max_{\theta} \sum_{S_k \in \Sigma^*} \mathcal{D}^{'}(f_{\theta}(c^*), S_k) P(S_k|\Sigma^*)
    \end{aligned}
\end{equation}

where \( \mathcal{D}^{'} \) is the distance metric between the target sequence \( f_{\theta}(c^*) \) and the sequence \( S_k \). Typically, we use the cross-entropy loss, defined as:

\begin{equation}
    \label{eq:RNA_design_loss}
    \mathcal{D}^{'}(f_{\theta}(c^*), S_k) = - \frac{1}{L} \sum_{i=1}^{L} \sum_{k=1}^{K} P(s_{ki}|f_{\theta}) \log P(s_{ki}|f_{\theta})
\end{equation}

\subsection{Flow Matching}

\subsubsection{Conditional Flow Matching}

Flow matching is an emerging generative method that, compared to diffusion-based approaches, eliminates the stochasticity in the noise-adding and denoising processes by transforming stochastic differential equations (SDEs) into ordinary differential equations (ODEs). Given a target data distribution $p_{data}(x) = p|_{t=t_{end}}(x)$ and a noise data distribution $p_{noise}(x) = p|_{t=0}(x)$, flow matching aims to learn a vector field $v(x,t)$ that transforms the noise distribution to the target distribution at a given time $t_{end}$, that is:
\begin{equation}
\begin{aligned}
  \frac{\partial p(x,t)}{\partial t} &= -\nabla \cdot (p(x,t)u(x,t)) \\
  p(x,0) &= p_{noise}(x) \\
  p(x,t_{end}) &= p_{data}(x)
\end{aligned}
\end{equation}

And the goal of flow matching is to minimize the difference between the predicted vector field $v(x,t)$ and the target vector field $u(x,t)$, that is:
\begin{equation}
\begin{aligned}
  \min_{\theta} \mathbf{E}_{x\sim p_{noise}(x),t\sim[0,t_{EndFor}]}[D(v_{\theta}(x,t),u(x,t))]
\end{aligned}
\end{equation}

\subsubsection{Dirichlet Flow Matching}

For RNA sequences, it is a common practice to output the probability distribution of bases at each position. This is precisely what Dirichlet flow matching accomplishes\citep{ref:dirichletfm}. By modeling the base probabilities as a Dirichlet distribution, this approach ensures that the generated sequences adhere to biologically plausible nucleotide distributions while maintaining the flexibility to explore diverse sequence spaces. Dirichlet flow matching leverages the properties of the Dirichlet distribution to smoothly transform an initial noise distribution into a target sequence distribution, enabling precise control over the generative process.

Considered a Dirichlet distribution with concentration parameter $\alpha = \{ \alpha_1, \alpha_2, \alpha_3, \alpha_4 \}$, the probability density function of the Dirichlet distribution is given by:
\begin{equation}
  Dir(\bf{x}; \alpha_1, \ldots \alpha_K) = \frac{1}{\mathcal{B}(\alpha_1, \ldots \alpha_K)}\prod_{i=1}^K x_i^{\alpha_i-1}
\end{equation}

Then we can define a conditional probability path with $t \in [0, \infty)$ as:
\begin{equation}\label{eq:probability-path}
  p_t(\bf{x} \mid \bf{x_1} = \bf{e_i}) = Dir(\bf{x}; \boldsymbol{\alpha} = \boldsymbol{1} + t\cdot\bf{e_i})
\end{equation}
and the target vector field is
\begin{equation}\label{eq:vectorfield}
    u_t(\bf{x} \mid \bf{x_1} = \bf{e_i}) = C(x_i, t) (\bf{e_i} -\bf{x})
\end{equation}
where $C(x_i, t)$ is a constant function of $x_i$ and $t$.

\begin{equation}
\label{eq:C_factor}
  C(x_i, t) =  - \tilde I_{x_i}(t+1, K-1)\frac{\mathcal{B}(t+1, K-1)}{(1-x_i)^{K-1} x_i^t}
\end{equation}
and
\begin{equation}
  \tilde I_x(a, b) = \frac{\partial }{\partial a}I_x(a, b)
\end{equation}

\subsection{RNACG}

\subsubsection{Training Goal} According to Eq \eqref{eq:RNA_design_function}, \eqref{eq:RNA_design_goal_2}, and \eqref{eq:RNA_design_loss}, we can define the RNACG as
\begin{equation}
P(S|f_{\theta})=f_{\theta}(x;C)
\end{equation}
where $S = \{A,C,G,U\}^{L}$ is the generated RNA sequence, $x$ is the initial input sampled from a given distribution, and $C$ is the conditional input. 

By tokenization and one-hot encoding, the RNA sequence can be represented as a matrix $S \in \mathbf{R}^{L \times K}$, where $L$ is the sequence length and K is the number of tokens, or,
\begin{equation}
S = \{s_i = \mathbf{e_k}\}^{L}
\end{equation}
where $s_i$ is the i-th token of $S$, and $\mathbf{e_k}$ is the one-hot encoding of the k-th class of token $s_i$.

The model is trained to minimize the difference between the generated sequence distribution and the target distribution, that is
\begin{equation}
\min_{\theta} \mathbf{E}_{x\sim p(x),\hat{S}\sim \Sigma^{*}(C)}[D(f_{\theta}(x;C),\hat{S})]
\end{equation}
where $D$ is a distance metric between the generated sequence $f_{\theta}(x;C)$ and the target sequence $\hat{S}$, and $p(x)$ and $\Sigma^*(C)$ are the initial distribution and the target distributions, respectively. As mentioned in Eq \eqref{eq:RNA_design_loss}, we choose cross-entropy loss as the distance metric and loss function.

\subsubsection{Model Architecture}

\begin{figure*}[h]
  \centering
  \includegraphics[width=0.9 \textwidth]{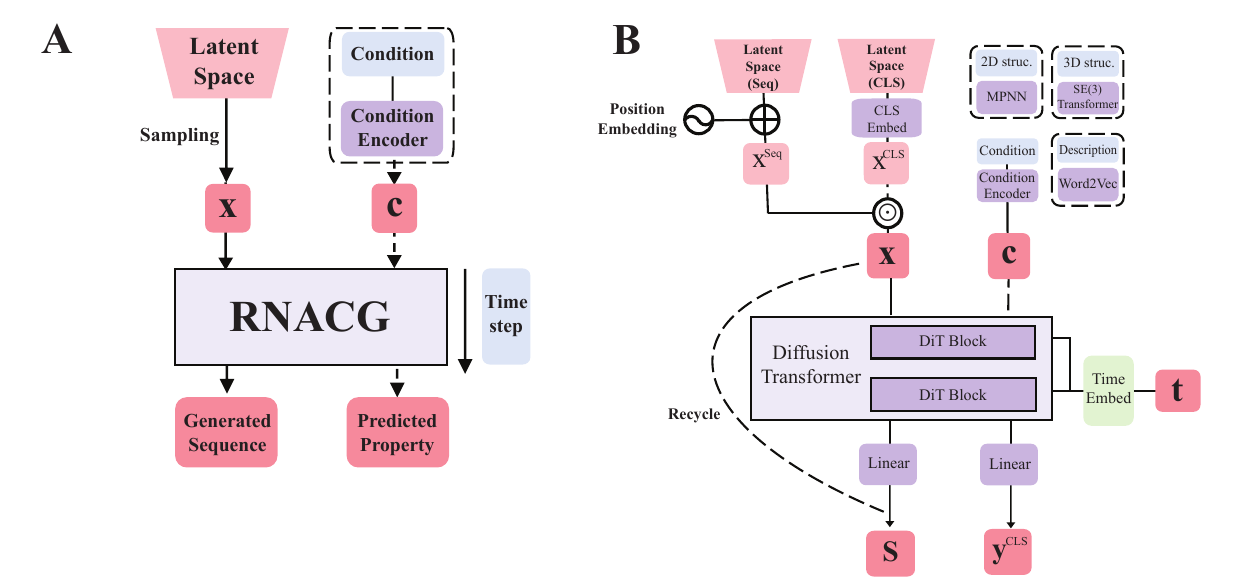} 
  \caption{Overview of the RNACG workflow. }
  \label{fig1}
\end{figure*}

The main body of RNACG comprises three components as shown in Figure\ref{fig1}: a random sampler and its corresponding Embedding layer, a multimodal Diffusion Transformer \citep[mm-DiT]{ref:dit, ref:sd3dit}, and a downstream Linear layer for sequence generation. In the case of conditional generation tasks, there is an additional part called Condition Encoder, which is a customized module based on the input conditions. Training and prediction can be performed using the CLS token for label-guided generation or prediction of input/output sequence properties.

Here, we can describe the DiT as follows:
$$
y_t = f^{DiT}_{\theta}(x,t;C) = f^{DiT}_{\theta}(x,t,c)
$$
where x is
$$
x = \textrm{Concat}(x^{cls},x^{seq})
$$
and c is
$$
c = \textrm{Condition Encoder}(C)
$$
And the predicte vector field is
$$
v^{seq}_{\theta} = \textrm{Linear}(y_{\theta}^{seq}),
v^{cls}_{\theta} = \textrm{Linear}(y_{\theta}^{cls})
$$

Noted, the [CLS] token can be used to classifier-guided generation and property prediction, and the vector field of [CLS] token is not neccessary to be predicted. We will clarify this in the following sections.

\subsubsection{Condition Encoder for Different Tasks}

For family-specific sequence generation, the conditional input is the Rfam family accession ID. We just use a single embedding layer to encode the family ID and merge it with the [CLS] token, not the condition encoder, that is, $x^{cls} = \textrm{Embedding}(\textrm{Rfam Accession ID})$ and $c = \emptyset$.

For 3D structure inverse folding, the conditional input is the 3D structure of the RNA sequence. Considered the success of ProteinMPNN in protein design, we use the same architecture as the Condition Encoder. We use the coordinates of the atoms O3', C3', C4', C5', O5', and P as the input features. It ensures no information leakage about the category of the nucleotides because the selected atoms are the same for all nucleotides.

Then we can define the input of Condition Encoder as
\begin{equation}
  \begin{aligned}
    c &= \textrm{MPNN}(\bf{r}_{O3'}, \bf{r}_{C3'}, \bf{r}_{C4'}, \bf{r}_{C5'}, \bf{r}_{O5'}, \bf{r}_{P}) \\
  \end{aligned}
\end{equation}

For UTR TE prediction, the conditional input is the species of the RNA sequence. We use a single embedding layer to encode the species ID and merge it with the [CLS] token, that is, $x^{cls} = \textrm{Embedding}(\textrm{Species ID})$ and $c = \emptyset$. 

Specifically, we use the vector field of the [CLS] token to predict the property of the sequence, that is, $v^{cls}_{\theta} = \textrm{Linear}(y_{\theta}^{cls})$. The sampling and generation process is the same as we have described in Eq \eqref{eq:normal sampling} and Eq \eqref{eq:loss of normal sampling}.

\subsubsection{Sampling and Training Strategy}
Normally, the initial $x$ can be sampled from a standard normal distribution, just as what we have done for $x^{cls}$ sometimes, that is $x^{cls} \sim \mathcal{N}(0,1)$. The target vector field is
\begin{equation}\label{eq:normal sampling}
\begin{aligned}
  u^{cls}_t &= x_t^{cls} - y^{cls} \\
  x_t^{cls} &\sim \mathcal{N}((1-t)x^{cls}_0+ty^{cls},\sigma^2)  
\end{aligned}
\end{equation}
and the loss function should be L2 loss, that is
\begin{equation}\label{eq:loss of normal sampling}
  \begin{aligned}
    \mathcal{L}^{cls}(\theta) = ||v^{cls}_{\theta}(x,t;C)-u^{cls}(x,t;C)||^2
  \end{aligned}
\end{equation}

However, using normal distribution to sample the sequence probability distribution is not suitable for RNA sequence data. So we use the Dirichlet distribution to sample the sequence probability distribution, that is

\begin{equation}
\begin{aligned}
  x^{seq} &\sim \mathbf{Dir}(\alpha) \\
  \alpha &= \mathbf{1} + t \cdot S
\end{aligned}
\end{equation}
and
\begin{equation}
  \mathbf{Dir}(x|\alpha) = \frac{\Gamma(\sum_{i=1}^{K}\alpha_i)}{\prod_{i=1}^{K}\Gamma(\alpha_i)}\prod_{i=1}^{K}x_i^{\alpha_i-1} = \frac{1}{B(\alpha)}\prod_{i=1}^{K}x_i^{\alpha_i-1}
\end{equation}
where $\alpha$ is the concentration parameter, $\Gamma(x)$ is the gamma function, and $B(\alpha)$ is the beta function. 

Then we can give the target vector field as
\begin{equation}
  \begin{aligned}
    u^{seq}_t &= x_t^{seq} - y^{seq} \\
    x_t^{seq} &\sim \mathbf{Dir}(\alpha = \mathbf{1} + t \cdot S)  
  \end{aligned}
\end{equation}

Specifically, during training, we observed that if uniform weights were applied to all positions, the model tended to converge to a state approximating direct argmax sampling. This behavior is intuitive because, during training, the intermediate time steps involve partially noised data, making direct argmax a natural local optimum. To address this, we introduced a weighted training strategy based on the discrepancy between input and target values: positions where the target and input values were closer were assigned smaller weights, while positions with larger deviations were given higher weights. This ensured that the model focused more on regions with greater discrepancies during training, that is

\[
\mathcal{L}_{\text{weighted}}(x_t, \text{output}, B) = \frac{1}{N} \sum_{i=1}^{N} w_i \cdot \text{CE}(x_t^{(i)}, \text{output}^{(i)})
\]

where \( N \) is the number of positions in the sequence, \( w_i \) is the weight for the \( i \)-th position, computed based on the discrepancy between the target and predicted values and \( \text{CE}(x_t^{(i)}, \text{output}^{(i)}) \) is the standard cross-entropy loss for the \( i \)-th position.

The weights \( w_i \) are normalized such that:

\[
w_i = \frac{w_i}{\sum_{j=1}^{N} w_j} \cdot N
\]

During the validation phase, however, we reverted to using a standard cross-entropy loss with equal weights to evaluate the model's performance. The detailed training algorithm and weighted loss function are provided in Algorithms \ref{alg:training} and \ref{alg:weighted_loss}, respectively. And the sequence generation algorithm is provided in Algorithm \ref{alg:generation}.

\begin{algorithm}
\caption{Training Algorithm}
\label{alg:training}
\begin{algorithmic}[1]
\Require Model $M$, optimizer $O$, loss function $\mathcal{L}$, training dataset $\mathcal{D}_{\text{train}}$, number of epochs $E$
\Ensure Trained model $M$

\State \textbf{Initialize:} Model parameters $\theta$, optimizer $O$, loss function $\mathcal{L}$

\For{epoch $= 1$ to $E$}
    \For{each batch $B \in \mathcal{D}_{\text{train}}$}
        \State $x_1 \gets \text{OneHotEncode}(B)$ \Comment{Encode input sequences}
        \State $x_t, t \gets \text{SampleConditionalPath}(x_1)$ \Comment{Ref Eq \eqref{eq:probability-path}}
        \State $\text{output} \gets M(x_t, t)$ 
        \State $\text{loss} \gets \mathcal{L}(x_t, \text{output}, B)$ 
        \State $\text{loss.backward()}$ 
        \State $O.\text{step()}$ 
    \EndFor
\EndFor

\State \textbf{Return:} Trained model $M$
\end{algorithmic}
\end{algorithm}

\begin{algorithm}
\caption{Weighted Cross-Entropy Loss}
\label{alg:weighted_loss}
\begin{algorithmic}[1]
\Require Predicted probabilities \( \text{output} \), target sequences \( S \), input sequences \( x_t \)
\Ensure Weighted loss \( \mathcal{L}_{\text{weighted}} \)

\State \textbf{Compute weights:}
\For{each position \( i \) in the sequence}
\State \( w_i \gets \text{CrossEntropy}(x_t^{(i)}, S^{(i)}) \) \Comment{Compute discrepancy as weight}
\EndFor
\State \( w \gets \text{Normalize}(w) \) \Comment{Normalize weights}

\State \textbf{Compute weighted loss:}
\State \( \mathcal{L}_{\text{weighted}} \gets 0 \)
\For{each position \( i \) in the sequence}
\State \( \mathcal{L}_{\text{weighted}} \gets \mathcal{L}_{\text{weighted}} + w_i \cdot \text{CrossEntropy}(\text{output}^{(i)}, S^{(i)}) \)
\EndFor
\State \( \mathcal{L}_{\text{weighted}} \gets \frac{\mathcal{L}_{\text{weighted}}}{N} \) \Comment{Average over sequence length}

\State \textbf{Return:} \( \mathcal{L}_{\text{weighted}} \)
\end{algorithmic}
\end{algorithm}

\begin{algorithm}
\caption{Sequence Generation Algorithm}
\label{alg:generation}
\begin{algorithmic}[1]
\Require Model $M$, input conditions $c$, attention mask $\text{attn\_mask}$, timesteps $t_{\text{span}}$
\Ensure Generated sequences $S_{\text{gen}}$

\State \textbf{Initialize:} $x_t \gets \textrm{Dir}(\textbf{1})$ \Comment{Initialize with noise or uniform distribution}

\For{each pair $(t_{start}, t_{end}) \in \text{zip}(t_{\text{span}}[:-1], t_{\text{span}}[1:])$}
  \State $\text{output} \gets M(x_t, s, c)$ 
  \State $c_{\text{factor}} \gets \text{ComputeCFactor}(x_t, s)$ \Comment{Ref: Eq \eqref{eq:C_factor}}
  \State $\text{flow\_probs} \gets \text{Softmax}(\text{output})$ 
  \State $\text{cond\_flow} \gets (\text{eye} - x_t \otimes \mathbf{1}) \cdot c_{\text{factor}}$ 
  \State $\text{flow} \gets (\text{flow\_probs} \otimes \text{cond\_flow}).\text{sum}(-1)$ 
  \State $x_t \gets x_t + \text{flow} \cdot (t_{end} - t_{start})$ 
\EndFor

\State $S_{\text{gen}} \gets \text{argmax}(x_t)$ \Comment{Convert final probabilities to sequences}
\State \textbf{Return:} $S_{\text{gen}}$
\end{algorithmic}
\end{algorithm}

\subsection{Evaluation Metrics}
We evaluate the family specificity of generated sequences using CM files from the Rfam database and the \texttt{cmsearch} program from Infernal\citep{ref:rfam}. For the 3D structure inverse folding task, we use sequence recovery rate, macro-F1 score, and micro-F1 score to assess model performance.

The sequence recovery rate is defined as:
\[
\text{Sequence Recovery Rate} = \frac{1}{L} \sum_{i=1}^{L} \mathbb{I}(y_i = \hat{y}_i),
\]
where \( y_i \) is the true nucleotide class (\( y_i \in \{A, C, G, U\} \)), \( \hat{y}_i \) is the predicted class, and \( \mathbb{I}(\cdot) \) is the indicator function.

For each class \( c \in \{A, C, G, U\} \), compute precision and recall:
\[
\text{Precision}_c = \frac{\text{TP}_c}{\text{TP}_c + \text{FP}_c}, \quad
\text{Recall}_c = \frac{\text{TP}_c}{\text{TP}_c + \text{FN}_c}.
\]
The F1-score for class \( c \) is:
\[
\text{F1}_c = 2 \cdot \frac{\text{Precision}_c \cdot \text{Recall}_c}{\text{Precision}_c + \text{Recall}_c}.
\]
The macro-averaged F1-score is:
\[
\text{F1-score}_{\text{macro}} = \frac{1}{4} \sum_{c \in \{A, C, G, U\}} \text{F1}_c.
\]

Compute global precision and recall by aggregating counts across all classes:
\[
\text{Precision}_{\text{micro}} = \frac{\sum_{c} \text{TP}_c}{\sum_{c} (\text{TP}_c + \text{FP}_c)}, \quad
\text{Recall}_{\text{micro}} = \frac{\sum_{c} \text{TP}_c}{\sum_{c} (\text{TP}_c + \text{FN}_c)}.
\]
The micro-averaged F1-score is:
\[
\text{F1-score}_{\text{micro}} = 2 \cdot \frac{\text{Precision}_{\text{micro}} \cdot \text{Recall}_{\text{micro}}}{\text{Precision}_{\text{micro}} + \text{Recall}_{\text{micro}}}.
\]

\section{Experiments}

\subsection{Dataset and Baselines}

Before formally introducing the performance of RNACG across different tasks, we need to clarify the dataset construction methods and the baseline models selected for each task. To facilitate the evaluation of generated sequence quality, we restricted the datasets to specific RNA families and used the covariance models (CM models) from Rfam to assess sequence specificity. Additionally, to enable a parallel comparison between classifier-guided generation (based on family classification information) and inverse folding generation (based on tertiary structures), we selected sequences from the Rfam.pdb file. These sequences are annotated by Rfam and have corresponding tertiary structure information in the PDB database—the same dataset used by Ribodiffusion for training and evaluation.

For the task of generating family-specific sequences guided by family classification information, we selected sequences from the Rfam.pdb file annotated with the families RF00001, RF00002, and RF00005. As baselines, we used the CMemit program from Infernal and RfamGen. Both baseline methods are based on CM models for sequence generation, with the key difference being that RfamGen optimizes the CM model construction process and uses a VAE model to adjust emission probabilities between different states.

For the task of sequence generation based on inverse folding of tertiary structures, we selected all sequences from Rfam.pdb that met the input format requirements (some structures were excluded due to missing coordinates of key atoms). For evaluation metrics such as sequence recovery rate and F1 score, we used RDesign and Ribodiffusion as baselines. We trained and compared our model using the training data provided by RDesign and Ribodiffusion. However, for family-specific evaluation, we only used Ribodiffusion as a baseline, as RDesign was not trained on the Rfam.pdb dataset. We avoided retraining RDesign to prevent potential performance misrepresentation due to inappropriate retraining procedures.

For the task of predicting 5'UTR translation efficiency, we used the training data published by UTR-LM \citep{ref:utrlm} to train our model.

\subsection{Classifier-Guided Generation of RNA Family-Specific Sequences}

\begin{figure*}[!h]\
  \centering
  \includegraphics[width=0.85\textwidth]{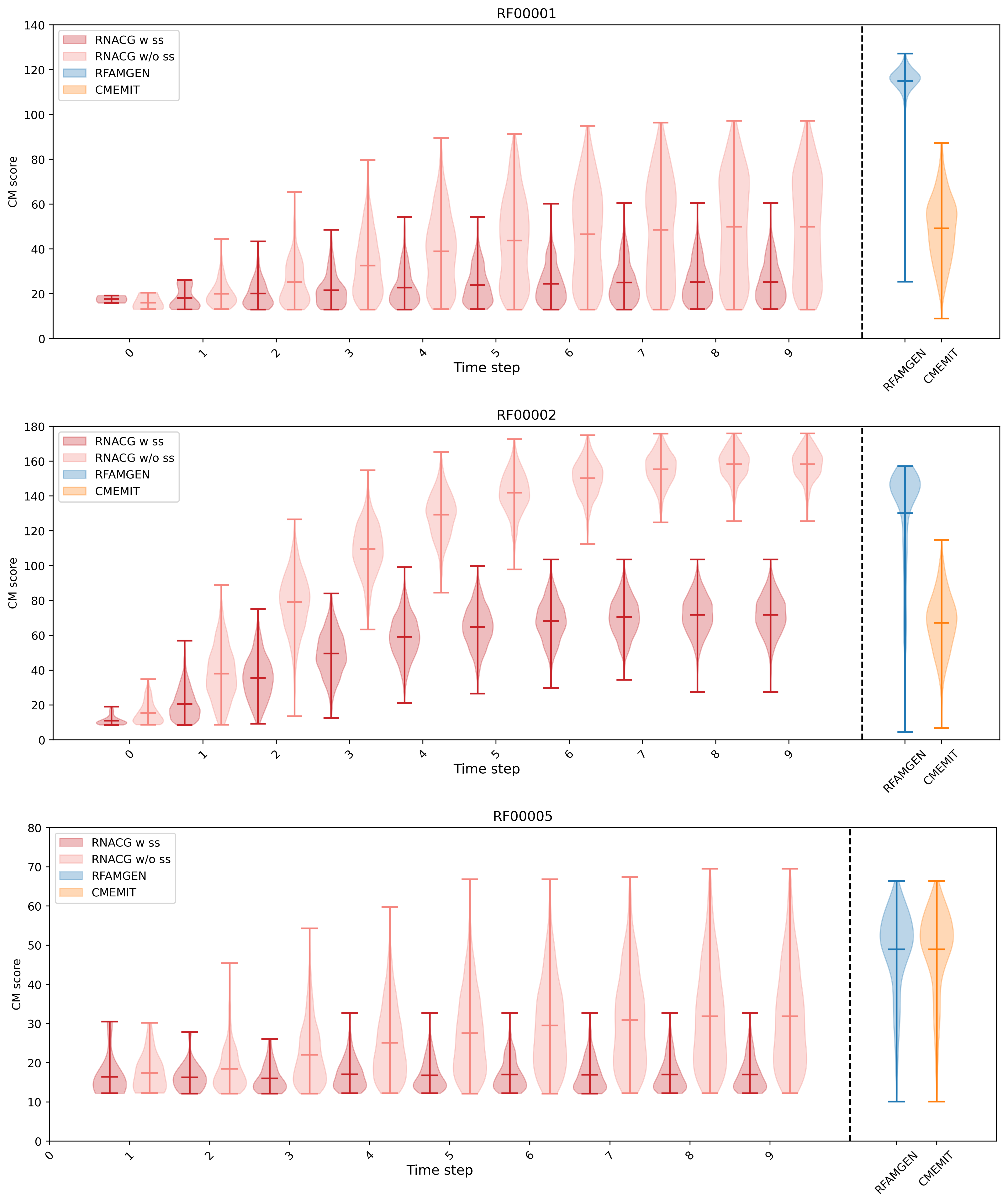}
  \caption{Evaluation of Sequence Generation Methods Across RNA Families.
  The figure compares the performance of RNACG (with and without secondary structure constraints), RfamGen, and cmemit across three RNA families (RF00001, RF00002, RF00005). Scores are plotted against time steps, demonstrating the improvement in sequence quality during generation. RNACG w/o ss (without secondary structure constraints) shows significant score improvements over time, while RNACG w/ ss (with secondary structure constraints) exhibits minimal improvement. Both RNACG and RfamGen generally outperform cmemit, highlighting the effectiveness of class embedding for family-specific sequence generation.}
  \label{fig:family}
\end{figure*}

The covariance model (CM) has been widely adopted as the gold standard for RNA family classification, forming the foundation of databases such as Rfam. Recent efforts, such as RfamGen (2024), have explored using Variational Autoencoders (VAEs) to refine emission probabilities in CM models for generating novel sequences. In this work, we propose a more flexible generation strategy that directly learns the sequence distribution of target RNA families. Our approach uses only the RNA family identifier (e.g., RF00001, RF00005) as input to guide sequence generation. To address the issue of the model exploiting shortcuts (e.g., selecting argmax under random noise-adding strategies), we introduce a weighted cross-entropy loss function. This function assigns smaller weights to regions with initial probability distributions close to the sampled sequences, encouraging the generation of diverse and biologically meaningful variations.

For comparison, we employed cmemit and RfamGen to generate sequences for the same RNA families and evaluated all generated sequences using the cmsearch program. To ensure fairness and feasibility, we did not use the entire set of candidate sequences from Rfam for training RNACG and RfamGen. Instead, we selected only those sequences with experimentally determined tertiary structures from Rfam for training. The specific sequences used are available in the supplementary materials or on GitHub.

Given that the number of sequences with tertiary structures is significantly smaller than the total number of sequences in Rfam, the performance of the retrained RfamGen model is naturally lower than the performance reported by its developers. This is expected due to the reduced training data size and the increased complexity of modeling sequences with tertiary structures.

As illustrated in the figure \ref{fig:family}, we investigated the impact of incorporating specific secondary structures (RNACG w/ ss) versus no secondary structure constraints (RNACG w/o ss) on sequence generation. The results reveal a striking contrast: sequences generated without secondary structure constraints exhibit a consistent and significant improvement in scores over time steps across all tested families (RF00001, RF00002, RF00005). In contrast, using predefined secondary structures for sequence correction resulted in minimal to no score improvement during generation.

Additionally, we compared the performance of RNACG (both with and without secondary structure constraints) against RfamGen and cmemit. While RNACG and RfamGen showed competitive performance, with each method outperforming the other in certain families, both methods generally surpassed the sequences generated by cmemit. This highlights the effectiveness of class embedding in achieving state-of-the-art performance for family-specific sequence generation, providing a robust and scalable foundation for future tasks.

\subsection{Inverse Folding on RNA 3D Structure}
Inverse folding, the task of designing sequences that fold into a specified tertiary structure, is a fundamental challenge in macromolecular design. While this approach has been widely applied in protein design, its development in RNA design has been more recent. Following the pioneering work of aRNAde, methods such as RDesign and Ribodiffusion have further explored RNA inverse folding. RDesign employs a network architecture and feature selection strategy similar to ProteinMPNN, while Ribodiffusion utilizes a diffusion-based approach to iteratively recover sequences.

In this work, we adopt a network architecture analogous to ProteinMPNN as our conditioning network. By integrating the embeddings generated by this conditioning network with the generative module of RNACG, we implement an RNA inverse folding pipeline based on flow matching. This approach leverages the strengths of both conditional modeling and flow-based generation to design sequences that accurately conform to target RNA tertiary structures.

\begin{table*}
  \begin{center}
    \begin{tabular}{cccccccccc}
      \toprule
      \textbf{Data Source} & \textbf{Method} & \multicolumn{2}{c}{\textbf{Short}} & \multicolumn{2}{c}{\textbf{Medium}} & \multicolumn{2}{c}{\textbf{Long}} & \multicolumn{2}{c}{\textbf{All}} \\
      & & Rec. & F1 & Rec. & F1 & Rec. & F1 & Rec. & F1 \\
      \midrule
      \multirow{4}{*}{RDesign} 
      & PiFold & 24.8 & - & 25.9 & - & 23.6 & - & 24.5 & - \\
      & RDesign & 37.2 & 38.2 & 44.9 & 42.5 & \underline{43.1} & 41.5 & 41.5 & 42.3 \\
      & UniIF & \underline{48.2} & - & \underline{49.7} & - & 37.3 & - & \underline{48.9} & - \\
      & \textbf{RNACG} & \textbf{50.9} & \textbf{49.3} & \textbf{54.0} & \textbf{54.1} & \textbf{51.5} & \textbf{51.3} & \textbf{52.3} & \textbf{52.3} \\
      \cline{1-10}
      \multirow{2}{*}{\shortstack{Ribodiffusion\\Seq0.8}}
      & Ribodiffusion & \textbf{52.0} & - & 60.0 & - & 58.9 & - & 57.3 & - \\
      & \textbf{RNACG} & 35.5 & 33.4 & \textbf{63.7} & 62.9 & \textbf{75.9} & 75.1 & \textbf{71.0} & 70.2 \\
      \cline{2-10}
      \multirow{2}{*}{\shortstack{Ribodiffusion\\Struct0.6}}
      & Ribodiffusion & \textbf{61.5} & - & \textbf{73.9} & - & 58.0 & - & 66.5 & - \\
      & \textbf{RNACG} & 29.9 & 28.8 & 60.1 & 59.1 & \textbf{72.1} & 71.4 & \textbf{68.1} & 67.4 \\
      \toprule
    \end{tabular}
    \caption{Performance metrics (\%) of different RNA 3D inverse folding methods. Bold numbers indicate the best performance, and underlined numbers indicate the second-best performance. A dash (-) indicates unreported values. Seq0.8 and Struct0.6 refer to the Ribodiffusion datasets with sequence identity thresholds of 0.8 and structural similarity thresholds (based on TMscore) of 0.6, respectively.}
    \label{tab:compare_with_stan_li}
  \end{center}
\end{table*}

\begin{figure*}[!h]
  \centering
  \includegraphics[width=0.85\textwidth]{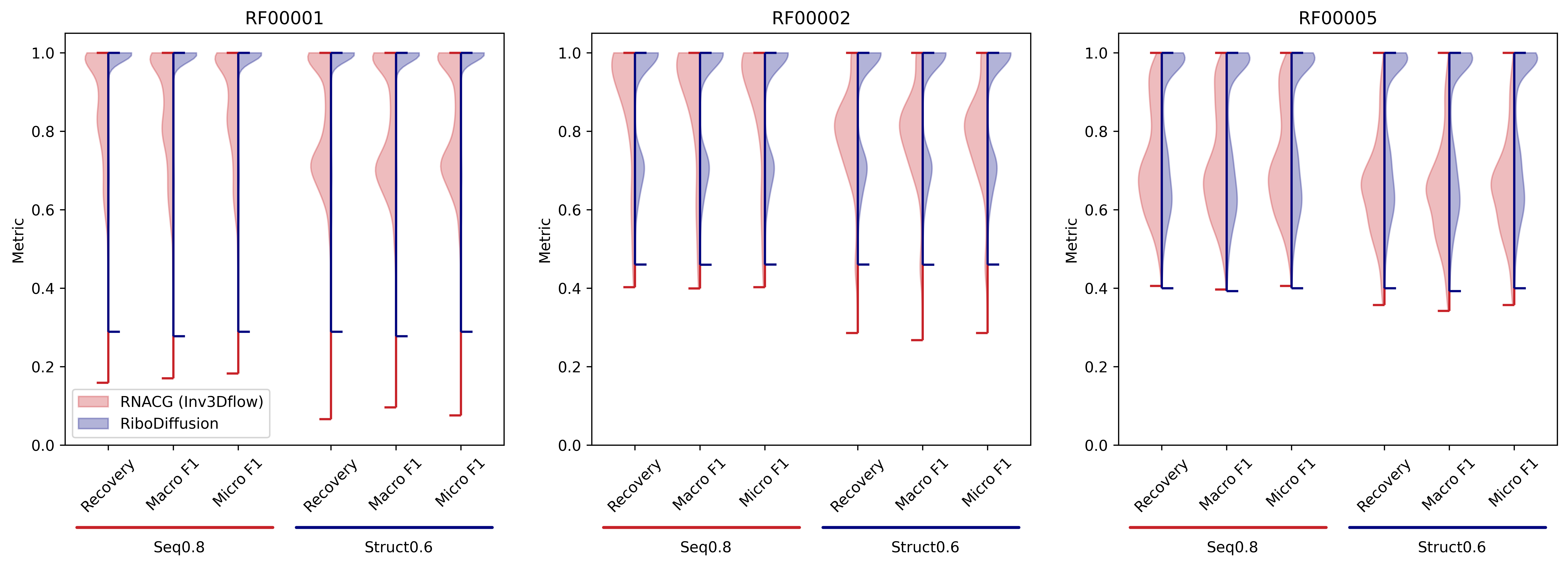}
  \caption{Evaluation of Inverse Folding Methods Across RNA Families on F1 Score.
  The figure compares the performance of RNACG and RiboDiffusion across three RNA families (RF00001, RF00002, RF00005). For each violin plot, the red half split represents the RNACG performance, while the blue half split represents the Ribodiffusion performance. And here shows Recovery Rate and F1 Score from left to right.}
  \label{fig:f1score}
\end{figure*}

We first trained our model on the training data provided by RDesign and Ribodiffusion, and the evaluation metrics are shown in the table \ref{tab:compare_with_stan_li}. From the table, it can be observed that our model outperforms the reported performance of PiFold, RDesign, and UniIF on the RDesign dataset. On the publicly available Ribodiffusion dataset, our model also surpasses the reported performance of Ribodiffusion on medium and long sequences. However, it is important to note that this comparison is not entirely fair. As mentioned earlier, some sequences were excluded from training due to missing atomic coordinates, which partially explains the relatively lower performance on short sequences. Furthermore, we did not include all the training data used by Ribodiffusion in our training process. For the reasons stated previously, we chose not to retrain Ribodiffusion. Therefore, we do not claim that RNACG achieves superior performance to Ribodiffusion; rather, we believe that the results demonstrate competitive performance.

Notably, as shown in table \ref{tab:compare_parameters}, our model has fewer than 5 million parameters, which is on the same order of magnitude as RDesign and significantly fewer than Ribodiffusion. This demonstrates the efficiency and effectiveness of our approach, as it achieves state-of-the-art performance with a compact model architecture.

\begin{table}[!h]
  \begin{center}
    \begin{tabular}{cccc}
      \toprule
      \textbf{Method} & \textbf{Parameters}\\
      \midrule
      RDesign & 1.3M \\
      Ribodiffusion & 52.2M\\
      RNACG & 4.5M \\
      \toprule
    \end{tabular}
    \caption{Comparison of model parameters}
    \label{tab:compare_parameters}
  \end{center}
\end{table}

\begin{figure*}[!h]
  \centering
  \includegraphics[width=0.85\textwidth]{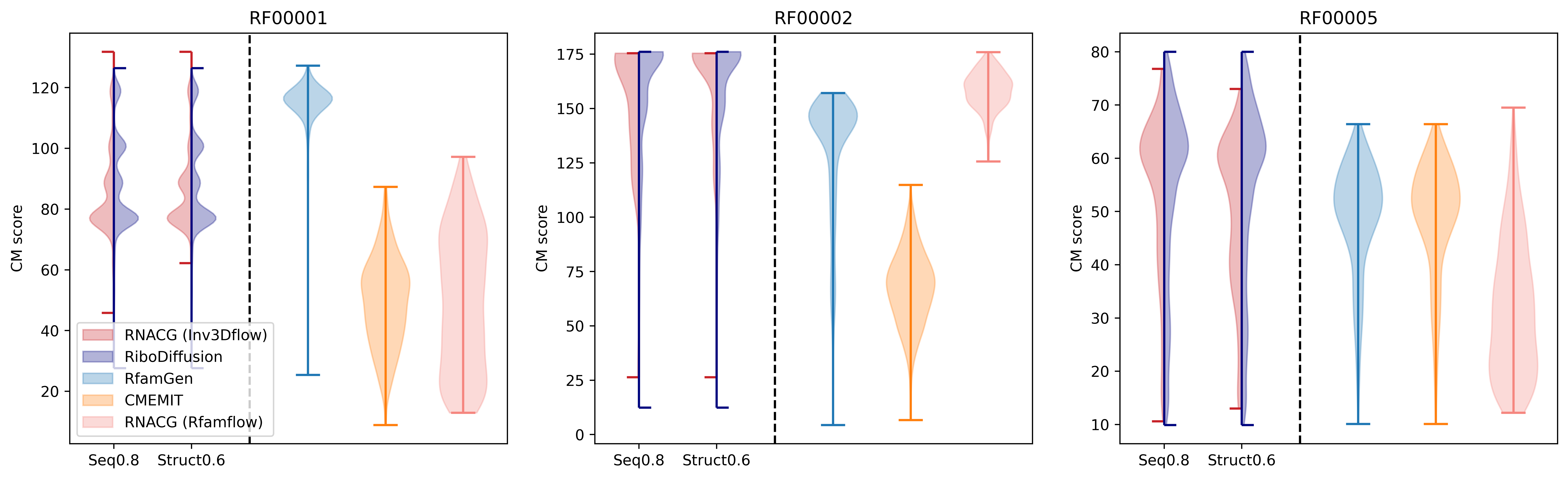}
  \caption{Evaluation of Sequence Generation Methods Across RNA families based cmsearch.
  The figure compares the performance of RNACG (Inv3Dflow, which based on 3D structure), RiboDiffusion, RfamGen, cmemit and RNACG (Rfamflow, which based on classifier-guided) across three RNA families (RF00001, RF00002, RF00005). }
  \label{fig:3dinv}
\end{figure*}

Then we evaluated the performance of both Ribodiffusion (using the model parameters provided by its developers) and RNACG on the test set. The evaluation metrics included sequence recovery rate, macro-F1 score, and micro-F1 score. Additionally, we also used cmsearch to assess the sequence specificity of the generated sequences based on tertiary structures from the Rfam database. As shown in figure \ref{fig:3dinv}, this comprehensive evaluation highlights the robustness and generalizability of both methods in the context of RNA inverse folding.

\subsection{5'UTR Translation Efficiency Prediction}


Finally, we explored using the CLS embedding from the RNACG model to predict the translation efficiency of 5'UTR sequences. While the model structurally still outputs generated 5'UTR sequences, this section focuses solely on the prediction of translation efficiency for input sequences, rather than their generation. The training process remains consistent with the earlier description. However, during prediction, we no longer start from a white noise distribution. Instead, we set the $\lambda$ value to 10, ensuring that the sequence recovery rate between the generated and target sequences remains above 0.95. We then employed a simple ResNet network as a post-processing module to predict translation efficiency based on the CLS embedding output.

\begin{table}[!h]
  \centering
  \begin{tabular}{lcccccc}
    \toprule
    \textbf{Celltype} & \textbf{Method} & \textbf{R\textsuperscript{2}} & \textbf{P. R} & \textbf{S. R} & \textbf{MAE} & \textbf{RMSE} \\
    \midrule
    \multirow{2}{*}{MUSCLE} 
    & RNACG & 0.434 & \textbf{0.708} & \textbf{0.690} & \textbf{0.617} & \textbf{0.829} \\
    & UTR-LM & \textbf{0.460} & 0.677 & 0.667 & 0.760 & 1.087 \\
    \hline
    \multirow{2}{*}{PC3} 
    & RNACG & \textbf{0.524} & \textbf{0.794} & \textbf{0.644} & 0.599 & 0.813 \\
    & UTR-LM & 0.469 & 0.685 & 0.646 & \textbf{0.509} & \textbf{0.702} \\
    \hline
    \multirow{2}{*}{HEK293} 
    & RNACG & \textbf{0.482} & \textbf{0.784} & \textbf{0.684} & 0.651 & 0.880 \\
    & UTR-LM & 0.409 & 0.639 & 0.603 & \textbf{0.572} & \textbf{0.855} \\
    \toprule
  \end{tabular}
  \caption{Performance metrics comparing UTR-LM and RNACG. Bold values indicate the best performance for each metric. P. R: Pearson correlation; S. R: Spearman correlation.}
  \label{tab:table4}
\end{table}

Our results shown as table \ref{tab:table4} demonstrate that RNACG achieves performance comparable to UTR-LM in the task of 5'UTR prediction. This highlights the potential of RNACG in controlling specific properties of generated sequences—such as the translation efficiency of UTR sequences—providing a promising direction for more fine-grained control over sequence generation.

Nonetheless, interpretive challenges persist. While our testing methodology employed actual values over sampling to ensure a precise mapping between outputs and true values, the actual generation process involves continuous probability inputs, diverging from the discrete one-hot encoding of real values. The interplay among input variables, predicted output values, and the resulting sequences must be meticulously considered throughout the generation process. Nonetheless, the precision in predicting output values is pivotal for explicit iterative optimization of sequences. Through successive iterations, we refine the output's probability distribution, progressively aligning it with the one-hot encoding ideal.

\section{Conclusion}

We introduce RNACG, a universal RNA sequence design method, demonstrating superior or competitive performance across multiple benchmarks. Our model's predictive capabilities enable iterative optimization of sequences, enhancing its applicability.

Our RNA 3D inverse folding tests confirmed that RNACG enriches sequence generation with valuable semantic insights. Pre-training on extensive RNA databases, exemplified by our 5'UTR translation efficiency predictions, is a promising strategy.

The model faces limitations, primarily in defining the RNA sequence sampling space. For example, encoding deviations from CM information can lead to numerous ineffective sequence designs. Our current token set, limited to [CLS] and ACGU with fixed-length outputs, restricts the generation scope. However, by shifting the focus from property to sequence distances, we've addressed the sparse sampling issue in training. This approach, however, introduces bias by directly altering the sequence sampling space rather than the property space.

In conclusion, RNACG offers a broadly applicable, customizable RNA sequence design tool. Its potential contributions to vaccine development and genetic engineering are expected to be substantiated by rigorous biological testing.

\bibliographystyle{abbrvnat}
\bibliography{aaai25}

\end{document}